\documentclass{jpconf}
\usepackage[paperheight=297mm,paperwidth=210mm,height=230mm,width=160mm,top=40mm,bottom=27mm,left=25mm,right=25mm]{geometry}
\usepackage{amsmath,amssymb,graphicx}
\begin{document}
\title{Yang-Mills connections valued on the octonionic algebra}
\author{A Restuccia$^{1,2}$ and J P Veiro$^3$}
\address{$^1$Universidad de Antofagasta, Departamento de F\'{\i}sica, Antofagasta, Chile}
\address{$^2$Universidad Sim\'on Bol\'{\i}var, Departamento de F\'{\i}sica, Apartado 89000, Caracas 1080-A, Venezuela}
\address{$^3$Universidad Sim\'on Bol\'{\i}var, Departamento de Matem\'aticas Puras y Aplicadas, Apartado 89000, Caracas 1080-A, Venezuela}
\ead{arestu@usb.ve and jpveiro@usb.ve}
\begin{abstract}
We consider a formulation of Yang-Mills theory where the gauge field is valued on an octonionic algebra and the gauge transformation is the group of automorphisms of it. We show, under mild assumptions, that the only possible gauge formulations are the usual $\mathfrak{su}(2)$ or $\mathfrak{u}(1)$ Yang-Mills theories.
\end{abstract}
\section{Introduction}
Yang-Mills theory with appropriate structure group is the basic ingredient in the description of the fundamental forces in nature. A notorious extension of Yang-Mills theory is Super Yang-Mills in 3, 4, 6, and 10 spacetime dimensions. It is well known the relation between the division algebras: the reals ${\mathbb R}$, complex ${\mathbb C}$, quaternion ${\mathbb H}$, the octonions ${\mathbb O}$, and Super Yang-Mills~\cite{KugoTownsend,Duff,ManogueSchray,BaezHuerta10,BaezHuerta11,Evans88,Evans95,BorstenDahanayakeDuffEbrahimRubens,AnastasiouBorstenDuffHughesNagy-SYM}. In particular octonions are relevant since the algebra $\mathfrak{sl}(2,{\mathbb O})$ is isomorphic to the $\mathfrak{so}(9,1)$ Lorentz algebra~\cite{Sudbery83,Sudbery84,SudberyBarton,Veiro,ManogueSchray,ManogueDray99,ManogueDray10}, the relativistic symmetry in string theory. It can also be used to describe representations of the Lorentz group in $D=11$ dimensional space-time where M-theory is formulated. The octonionic algebra was relevant to obtain Yang-Mills instanton solutions in seven and eight dimensions~\cite{GunaydinNicolai,FubiniNicolai}. Recently the relation between Yang-Mills, division algebras, and triality has been emphasized~\cite{AnastasiouBorstenDuffHughesNagy-SYM,BorstenDuffHughesNagy}, in particular an octonionic formulation of the $N=1$ Supersymmetry algebra in $D=11$ space-time dimensions has been obtained~\cite{AnastasiouBorstenDuffHughesNagy-MTheory}.
\\

In all these interesting formulations the gauge field is always valued on a Lie algebra. We will consider in this work the construction of Yang-Mills theory where the gauge field is valued on the algebra of the octonions. In order to do so we have to define the corresponding gauge transformations and consider its action on the curvature of the connection one-form. The formulation we will consider is based on a proposition introduced by Okubo~\cite{Okubo}.
\\

The main result of this work is that under mild assumptions: \eqref{cond-i}, \eqref{cond-ii}, and~\eqref{cond-iii} in Section~\ref{sec3}, the only possible gauge formulation, with the gauge field valued on the octonionic algebra, of a Yang-Mills theory are the ones on which the connection is valued on an associative subalgebra of the octonions.
\section{The Yang-Mills formulation on a non-associative algebra}
\label{sec3}
Given a Lie algebra $L$, let us denote by $\Lambda_p$, for $p>0$, the space of all $L$-valued $p$-forms and $\Lambda_0=L$. The boundary operator, ${\rm d}:\:\Lambda_p\to\Lambda_{p+1}$, satisfies the usual properties:
\begin{enumerate}
\item
${\rm d}{\rm d}=0$,
\item
${\rm d}(\omega_p\wedge\omega_q)={\rm d}\omega_p\wedge\omega_q+(-1)^p\omega_p\wedge{\rm d}\omega_q$ where $\omega_p\in\Lambda_p$ and $\omega_q\in\Lambda_q$ are $L$-valued $p$-forms and $q$-forms respectively.
\end{enumerate}
Consider $\omega$ to be an $L$-valued one-form which we shall write
\[
\omega=\sum_{\mu=1}^mA_\mu{\rm d}x^\mu,\]
where the ${\rm d}x^\mu$ are anti-commutative one-forms, and the curvature is then given by the $L$-valued two-form $R={\rm d}\omega+\omega\wedge\omega$ which can be expressed in terms of $F_{\mu\nu}$ via
\[
R=\sum_{1\leq\mu<\nu\leq m}F_{\mu\nu}{\rm d}x^\mu\wedge{\rm d}x^\nu=\frac{1}{2}\sum_{\mu,\nu=1}^m F_{\mu\nu}{\rm d}x^\mu\wedge{\rm d}x^\nu.
\]
The gauge transformations are given by
\begin{equation}\label{gauge-w}
\omega\longrightarrow U^{-1}\omega U+U^{-1}{\rm d}U
\end{equation}
and they are such that the curvature transforms covariantly by
\begin{equation}\label{gauge-R}
R\longrightarrow U^{-1}RU.
\end{equation}
Given $A\in L$ write $g_U(A)=U^{-1}AU$, they are automorphisms of the Lie algebra. It is clear that $g_U\in GL(L)$ and it can also be extended to $\Lambda_p$ in a natural way for which it can simply be considered $g_U\in GL(\Lambda_p)$. If we write $\xi=U^{-1}{\rm d}U$ then $\xi$ satisfies the Maurer-Cartan relation
\[
{\rm d}\xi+\xi\wedge\xi=0.
\]
In this manner, Equations~\eqref{gauge-w} and~\eqref{gauge-R} can be expressed as
\begin{eqnarray*}
\omega&\longrightarrow&g_U(\omega)+\xi,\\
R&\longrightarrow&g_U(R).
\end{eqnarray*}

Consider now the octonion algebra ${\mathbb O}$. In~\cite{Okubo} Okubo proposed a generalization of Yang-Mills theory to non-associative algebras. Okubo's main idea for constructing a gauge theory based on a non-associative algebra $K$ is to use the group of automorphisms over $K$ as the group of gauge transformations. Let $\omega$ now be a one-form with coefficients valued over the non-associative algebra ${\mathbb O}$. To be precise,
\[
\omega=\sum_{\mu=1}^m\omega_\mu{\rm d}x^\mu
\]
with $\omega_\mu\in {\mathbb O}$. Analogously, the curvature two-form valued over the non-associative algebra ${\mathbb O}$ is given by
\[
R={\rm d}\omega+\omega\wedge\omega.
\]

For some $g\in GL({\mathbb O})$, and some $\xi\in\Lambda_1({\mathbb O})$, consider the gauge transformation
\begin{equation}\label{gauge-transf}
\omega\longrightarrow \omega^\prime=g(\omega)+\xi.
\end{equation}
In order that $R$ transforms covariantly as
\begin{equation}\label{transf-cov}
R\longrightarrow g(R)={\rm d}\omega^\prime+\omega^\prime\wedge\omega^\prime
\end{equation}
it is sufficient for $g\in GL(\Lambda_1)$ and $\xi\in\Lambda_1({\mathbb O})$ to satisfy the following conditions~\cite{Okubo}:
\begin{eqnarray}
&&g(\omega\wedge\omega)=g(\omega)\wedge g(\omega)\label{cond-i},\\
&&{\rm d}\xi+\xi\wedge\xi=0\label{cond-ii},\\
&&{\rm d}(g(\omega))=g({\rm d}\omega)-\xi\wedge g(\omega)-g(\omega)\wedge\xi\label{cond-iii}.
\end{eqnarray}
We notice that these conditions are satisfied for a Lie algebra Yang-Mills theory, however it is not known if they can be satisfied for an octonionic valued gauged field theory. The condition shown in Equation~\eqref{cond-i} is satisfied whenever $g$ is an automorphism over ${\mathbb O}$, that is $g\in{\rm Aut}({\mathbb O})$ which is the group $G_2$. In order to deal with Equations~\eqref{cond-ii} and~\eqref{cond-iii} it is convenient to translate them from the Lie group ${\rm Aut}({\mathbb O})$ to the Lie algebra $\mathfrak{Der}({\mathbb O})$. Given an infinitesimal parameter $\epsilon$ and a derivation $D\in\mathfrak{Der}({\mathbb O})$, we can always write
\[
g={\rm exp}(\epsilon D)={\rm id}+\epsilon D+O(\epsilon^2)
\]
where ${\rm id}$ is the identity transformation and
\[
\xi=\epsilon{\rm d}k+O(\epsilon^2)
\]
for some $k\in {\mathbb O}$. Since the boundary operator is such that $dd=0$, Equation~\eqref{cond-ii} holds automatically and Equation~\eqref{cond-iii} is equivalent to
\begin{equation}\label{cond-okubo}
{\rm d}(D(\omega))=D({\rm d}\omega)-{\rm d}k\wedge\omega-\omega\wedge{\rm d}k.
\end{equation}
Following Okubo's assumptions, \eqref{cond-i}, \eqref{cond-ii}, and \eqref{cond-iii}, it suffices to find pairs $(D,k)$ of elements in a subalgebra of $\mathfrak{Der}({\mathbb O})\times {\mathbb O}$ that satisfy Equation~\eqref{cond-okubo} for all $\omega\in\Lambda_1({\mathbb O})$ so that the gauge transformation~\eqref{gauge-transf} induces the covariant transformation~\eqref{transf-cov} as desired. This is the first step in the construction of the Yang-Mills theory. One should finally provide the action for the theory. Although \eqref{cond-i}, \eqref{cond-ii}, and \eqref{cond-iii} are mild assumptions, the non-associative context imposes very restrictive conditions on the gauge theory.
\section{The derivation algebra}
\label{sec4}
\begin{figure}[th]
\begin{center}
\includegraphics[width=4.8cm]{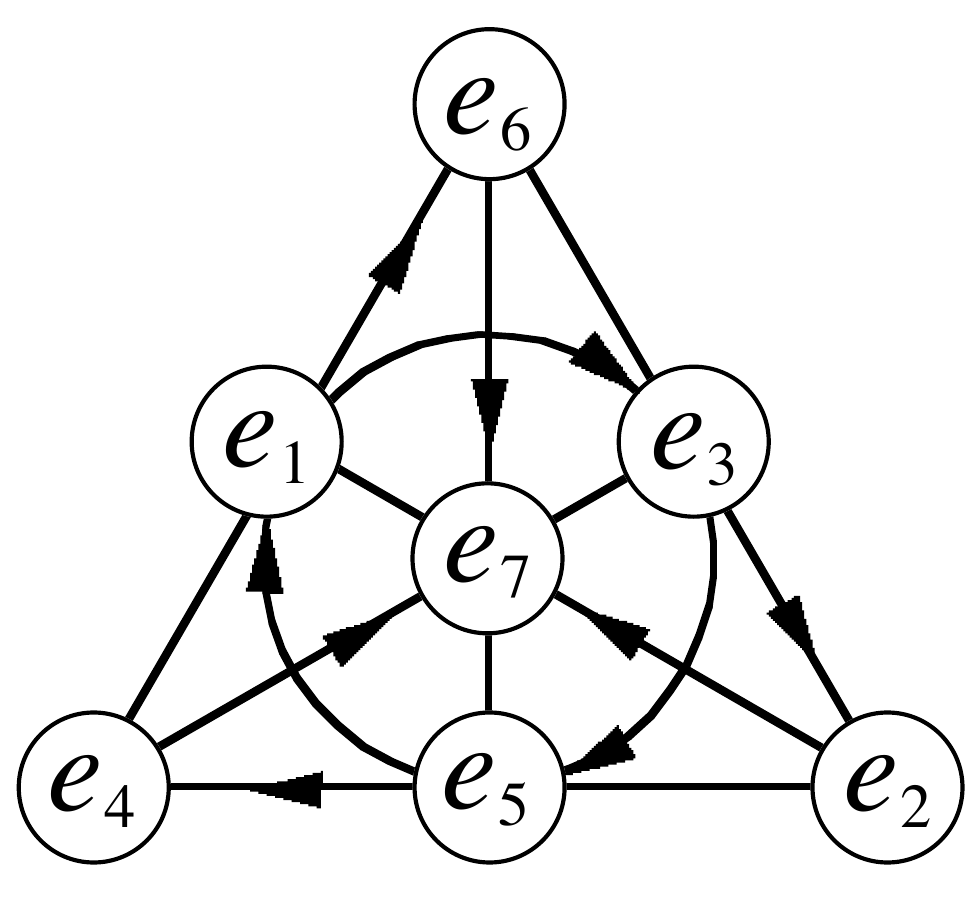}
\caption{Fano plane representing the multiplication table for the octonions.}
\label{fano}
\end{center}
\end{figure}

We shall consider a basis for the octonions where the multiplication of the imaginary elements is represented in the Fano plane shown in Figure~\ref{fano}.
The identity element $e_0$ is left outside because it commutes and associates with any other element. The diagram is read in the following way: the result of the product of any two elements is the only other element laying on the line that passes through the first two elements, and the sign is determined by the arrows (for example, $e_5e_2=-e_4$).
\\

Recalling Cartan's work~\cite{Cartan} on the classification of simple Lie groups, $G_2$, the smallest exceptional Lie group, is the group of automorphisms over the octonions. Hence $G_2={\rm Aut}({\mathbb O})$. The tangent space to a group of automotphisms is an algebra of derivations. Therefore, the Lie algebra $\mathfrak{g}_2$ of the Lie group $G_2$ is $\mathfrak{Der}({\mathbb O})$. Given two octonions, $a$ and $b$, the map
\begin{eqnarray*}
D_{a,b}(x)&=&\big([L_a,L_b]+[L_a,R_b]+[R_a,R_b]\big)(x)\\
&=&\frac{1}{2}\big([[a,x],b]+[a,[b,x]]+[[a,b],x]\big)\\
&=&[[a,b],x]-3[a,b,x],
\end{eqnarray*}
where $L_a(x)=ax$, $R_a(x)=xa$, the bracket with two entries is the commutator $[a,b]=ab-ba$, and the bracket with three entries is the associator $[a,b,x]=(ab)x-a(bx)$ acting over an octonion $x$, is a derivation. They clearly satisfy the Leibniz rule
\[
D_{a,b}(xy)=(D_{a,b}(x))y+x(D_{a,b}(y)),
\]
the Generalized Jacobi Identity
\[
D_{a,b}\left(D_{c,d}\right)=D_{D_{a,b}(c),d}+D_{c,D_{a,b}(d)}+D_{c,d}\left(D_{a,b}\right),
\]
and they are anti-symmetric in the sense that $D_{a,b}=-D_{b,a}$. For each index $p\in\{1,\ldots,7\}$, consider the three pairs of indices $(i,j)$, $(r,s)$, and $(u,v)$ such that $e_p=e_ie_j=e_re_s=e_ue_v$, then
\[
D_{e_i,e_j}+D_{e_r,e_s}+D_{e_u,e_v}=0.
\]
A basis for $\mathfrak{g}_2$ can be given with 14 of these transformations grouped in seven pairs, each of them constituted by $D_{e_i,e_j}$ and $D_{e_r,e_s}$ where $e_ie_j=e_re_s=e_p$ for a different $p$ in every one.
\section{The analysis of the Okubo restrictions}
\label{sec6}
We are interested in solving Equation~\eqref{cond-okubo} for the non-associative algebra of the octonions. It suffices to find pairs of elements $(D,k)$ in a subalgebra of $\mathfrak{Der}({\mathbb O})\times{\mathbb O}$ that satisfy Equation~\eqref{cond-okubo} for all $\omega\in\Lambda_1({\mathbb O})$. In what follows we will show that Okubo's formulation, for the case of the octonions being the non-associative algebra over which the geometric objects shall be valued, fails to extend the usual Yang-Mills theory formulated over Lie algebras.\\

Given two one-forms valued over the octonions, $\rho$ and $\omega$, they may be interpreted as one-forms with octonionic coefficients
\[
\rho=\rho_1{\rm d}x^1+\cdots+\rho_n{\rm d}x^n=\left(\sum_{i=0}^7\rho_1^ie_i\right){\rm d}x^1+\cdots+\left(\sum_{i=0}^7\rho_n^ie_i\right){\rm d}x^n
\]
and also as octonions with coefficients that are one-forms
\[
\rho=\rho^0e_0+\cdots+\rho^7e_7=\left(\sum_{i=1}^n\rho^0_i{\rm d}x^i\right)e_0+\cdots+\left(\sum_{i=1}^n\rho^7_i{\rm d}x^i\right)e_7
\]
simply by interchanging the summation order in the general expression. The product of one-forms valued over the octonions is given by
\[
\rho\wedge\omega=\sum_{1\leq i<j\leq n}\big(\rho_i\omega_j-\rho_j\omega_i\big){\rm d}x^i\wedge{\rm d}x^j
\]
which is a two-form valued over the octonions. Seen as an octonion, its real part is not necessarily equal to zero. Nonetheless, both $\omega\wedge\omega$ and $\rho\wedge\omega+\omega\wedge\rho$ do possess real part equal to zero. To be precise,
\[
\omega\wedge\omega=\sum_{1\leq i<j\leq n}\left[\omega_i,\omega_j\right]{\rm d}x^i\wedge{\rm d}x^j=\frac{1}{2}\sum_{i,j=1}^n\left[\omega_i,\omega_j\right]{\rm d}x^i\wedge{\rm d}x^j
\]
and
\begin{eqnarray*}
\rho\wedge\omega+\omega\wedge\rho&=&\sum_{1\leq i<j\leq n}\Big(\left[\rho_i,\omega_j\right]+\left[\omega_i,\rho_j\right]\Big){\rm d}x^i\wedge{\rm d}x^j\\
&=&\frac{1}{2}\sum_{i,j=1}^n\Big(\left[\rho_i,\omega_j\right]+\left[\omega_i,\rho_j\right]\Big){\rm d}x^i\wedge{\rm d}x^j\\
&=&\sum_{i,j=1}^n\left[\rho_i,\omega_j\right]{\rm d}x^i\wedge{\rm d}x^j.
\end{eqnarray*}
On the other hand, considering ${\cal G}\in\mathfrak{g}_2$ and using the fact that ${\rm d}^2=0$, we are able to calculate
\begin{eqnarray*}
{\rm d}\big({\cal G}(\omega)\big)-{\cal G}({\rm d}\omega)&=&\sum_{i=1}^n\Big({\rm d}\big({\cal G}(\omega_i)\big)\wedge{\rm d}x^i\Big)-{\cal G}\left(\sum_{i=1}^n{\rm d}\omega_i\wedge{\rm d}x^i\right)\\
&=&\sum_{i=1}^n\left({\rm d}\left(\sum_{j=1}^7\omega_i^j{\cal G}(e_j)\right)\wedge{\rm d}x^i
-\left(\sum_{j=1}^7{\rm d}\omega_i^j{\cal G}(e_j)\right)\wedge{\rm d}x^i\right)\\
&=&\sum_{i=1}^n\left(\sum_{j=1}^7\omega_i^j{\rm d}{\cal G}(e_j)\right)\wedge{\rm d}x_i
\end{eqnarray*}
noticing that ${\cal G}(e_0)=0$. Since ${\cal G}$ is a derivation over the octonions, it can be written as a linear combination over a basis of  $\mathfrak{g}_2$. The transformation ${\rm d}{\cal G}$ consists on applying the exterior derivative to the coefficients that determine ${\cal G}$. For example, if we write ${\cal G}={\displaystyle\sum_{i,j=1}^7}\lambda_{ij}D_{e_i,e_j}$ then
\begin{eqnarray*}
{\rm d}{\cal G}&=&{\rm d}\left(\sum_{i,j=1}^7\lambda_{ij}D_{e_i,e_j}\right)=\sum_{i,j=1}^7{\rm d}\lambda_{ij}D_{e_i,e_j}=\sum_{i,j=1}^7\left(\sum_{k=1}^n\partial_k\lambda_{ij}{\rm d}x^k\right)D_{e_i,e_j}\\
&=&\sum_{k=1}^n\partial_k{\cal G}{\rm d}x^k
\end{eqnarray*}
where $\partial_k{\cal G}={\displaystyle\sum_{i,j=1}^7}\partial_k\lambda_{ij}D_{e_i,e_j}$. Therefore,
\begin{eqnarray*}
{\rm d}\big({\cal G}(\omega)\big)-{\cal G}({\rm d}\omega)
&=&\sum_{i,j=1}^n\Big(
\omega_j^1\partial_i{\cal G}(e_1)+\cdots+\omega_j^7\partial_i{\cal G}(e_7)
\Big){\rm d}x^i\wedge{\rm d}x^j
\\
&=&\sum_{i,j=1}^n\Big(
\big(\partial_i{\cal G}\big)(\omega_j)
\Big){\rm d}x^i\wedge{\rm d}x^j.
\end{eqnarray*}
Finally, in order to solve the equation
\[
{\rm d}\big({\cal G}(\omega)\big)-{\cal G}({\rm d}\omega)=\rho\wedge\omega+\omega\wedge\rho
\]
where ${\cal G}\in\mathfrak{g}_2$ and $\rho=-{\rm d}k$ is a one-form valued over the octonions, it is necessary that ${\cal G}$ and $\rho$ satisfy the following relation, for any $\omega\in\Lambda_1({\mathbb O})$,
\begin{equation}\label{g2comocorchete}
\big(\partial_i{\cal G}\big)(\omega_j)=\left[\rho_i,\omega_j\right]
\end{equation}
for all indices $i,j\in\{1,\ldots,n\}$ such that $i\neq j$. Nonetheless, Equation~\eqref{g2comocorchete} only possesses non-trivial solutions, for the case of the octonions, when the expression is considered inside an associative subalgebra. To understand this situation, notice that $\partial_i{\cal G}$ is still a vector in $\mathfrak{g}_2$, thus a derivation and must satisfy the Leibniz rule. Therefore, if $a$ and $b$ are two octonions then
\begin{equation}\label{der-ab}
\partial_i{\cal G}(ab)=\big(\partial_i{\cal G}(a)\big)b+a\big(\partial_i{\cal G}(b)\big)
\end{equation}
and on the other hand
\begin{equation}\label{bracket-ab}
[\rho_i,ab]=[\rho_i,a]b+a[\rho_i,b]+3[\rho_i,a,b]
\end{equation}
by straightforward calculation. If it were possible that $\big(\partial_i{\cal G}\big)(\,\cdot\,)=\left[\rho_i,\,\cdot\,\right]$ then subtracting Equations~\eqref{der-ab} and~\eqref{bracket-ab} yields $[\rho_i,a,b]=0$. Since the octonion $\omega_i$ can always be written as a product of two octonions, Equation~\eqref{g2comocorchete} does not hold in general.
\\

Then, finding solutions to Equation~\eqref{g2comocorchete} is equivalent to answer the following question: When is it true that the transformation induced by the commutator bracket represents a transformation in $\mathfrak{g}_2$? The answer to this question is that it happens only when all the octonions into consideration belong to an associative subspace; that is, when the octonions involved belong to a subalgebra isomorphic to the quaternions. Artin's Theorem establishes that any subalgebra of the octonions generated by only two elements is associative; this statement is formulated in a more general context of alternative algebras~\cite[p. 29]{Schafer}. Returning to the question of wheter the commutator bracket is a derivation, given three octonions, $a$, $b$, and $c$, then
\[
[c,ab]=[c,a]b+a[c,b]+3[c,a,b]
\]
showing that $[c,ab]$ is equal to $[c,a]b+a[c,b]$ if and only if $[c,a,b]=0$. That is, the bracket on the right hand side of~\eqref{g2comocorchete} is a derivation only for an associative subalgebra of the octonions. This explains why it is impossible to find non-trivial solutions for Equation~\eqref{cond-okubo} when the non-associative algebra into consideration is the octonions. In this sense, the only formulation for a Yang-Mills theory on an octonionic algebra, under assumptions \eqref{cond-i}, \eqref{cond-ii}, and \eqref{cond-iii}, is the usual Yang-Mills theory for the algebras $\mathfrak{su}(2)$ and $\mathfrak{u}(1)$, since both the imaginary part of the quaternions and their algebra of derivations can be identified with $\mathfrak{su}(2)$ and it is in that context where the terms ${\rm d}(D(\omega))-D({\rm d}\omega)$ and $-{\rm d}k\wedge\omega-\omega\wedge{\rm d}k$ are compatible.
\section{Conclusions}
It was known since the work of Okubo that the conditions \eqref{cond-i}, \eqref{cond-ii}, and \eqref{cond-iii} for a non-associative algebra imply that the curvature two-form transforms in a natural way as an automorphism of the non-associative algebra in analogy with the Lie algebra case (in this case the three conditions are satisfied). We consider the construction of a Yang-Mills theory where the connection one-form is valued on the octonion algebra, in distinction to the construction where the structure group is a Lie group. We proved that the conditions \eqref{cond-i}, \eqref{cond-ii}, and \eqref{cond-iii} are not simultaneously satisfied for the octonion algebra. However they are satisfied for the associative subalgebras of the octonions: the quaternions and the complex numbers. They are associated to the $\mathfrak{su}(2)$ and $\mathfrak{u}(1)$ Yang-Mills theories respectively.
\section*{Acknowledgments}
A.R. is partially supported by Project Fondecyt 1121103.
\section*{References}

\end{document}